\documentclass[conference]{IEEEtran}
\IEEEoverridecommandlockouts

\usepackage{longtable}

\usepackage{float} 
\usepackage{adjustbox}
\usepackage{cite}
\usepackage{amsmath,amssymb,amsfonts}
\usepackage{algorithmic}
\usepackage{graphicx}
\usepackage{textcomp}
\usepackage{xcolor}

\usepackage{bm}

\usepackage{array}
\usepackage{stfloats}% <-- added

\def\BibTeX{{\rm B\kern-.05em{\sc i\kern-.025em b}\kern-.08em
    T\kern-.1667em\lower.7ex\hbox{E}\kern-.125emX}}

\begin{document}
\bstctlcite{IEEEexample:BSTcontrol}

\title{Deep Learning-based Power Control for Cell-Free Massive MIMO Networks\\
\thanks{This work was supported by the Academy of Finland 6Genesis Flagship (grant no. 318927).}
}

\author{
\IEEEauthorblockN{Nuwanthika~Rajapaksha, K. B. Shashika Manosha, Nandana~Rajatheva,  and Matti Latva-aho}
\IEEEauthorblockA{Centre for Wireless Communications, University of Oulu, Finland\\ 
E-mail: \{nuwanthika.rajapaksha, nandana.rajatheva, matti.latva-aho\}@oulu.fi, manoshadt@gmail.com}
}

\maketitle

\begin{abstract}

A deep learning (DL)-based power control algorithm that solves the max-min user fairness problem in a cell-free massive multiple-input multiple-output (MIMO) system is proposed. Max-min rate optimization problem in a cell-free massive MIMO uplink setup is formulated, where user power allocations are optimized in order to maximize the minimum user rate. Instead of modeling the problem using mathematical optimization theory, and solving it with iterative algorithms, our proposed solution approach is using DL. Specifically, we model a deep neural network (DNN) and train it in an unsupervised manner to learn the optimum user power allocations which maximize the minimum user rate. This novel unsupervised learning-based approach does not require optimal power allocations to be known during model training as in previously used supervised learning techniques, hence it has a simpler and flexible model training stage. Numerical results show that the proposed DNN achieves a performance-complexity trade-off with around 400 times faster implementation and comparable performance to the optimization-based algorithm. An online learning stage is also introduced, which results in near-optimal performance with 4-6 times faster processing.

\end{abstract}


\section{Introduction}

Massive MIMO, where a base station with a large number of antennas simultaneously serves many users, has become a key technology in fifth generation (5G) networks due to their high throughput and reliability \cite{6736746_5G, 5595728_marzetta, 6375940_scaling_mimo}. Cell-free massive MIMO combines both massive MIMO and distributed MIMO and has the potential of providing uniformly good throughput for all users \cite{7227028_cellfree_mimo_larson, 7827017_cellfree_vs_smallcells, 7917284_cellfree_precoding_power}. In cell-free massive MIMO, a large number of distributed access points (APs) serve a much smaller number of users distributed over a wide area, and there are no cells or cell-boundaries \cite{7227028_cellfree_mimo_larson}. All the APs are connected to a central processing unit (CPU) via the backhaul network, and they coherently cooperate to serve all users using same time-frequency resources via time-division
duplex (TDD) \cite{7827017_cellfree_vs_smallcells}. Different scalable cell-free massive MIMO architectures and receiver/combiner schemes for uplink/downlink signal processing are studied in \cite{8845768_emil_competitivecellfree, 9064545_scalable_cellfree, 7827017_cellfree_vs_smallcells, 7917284_cellfree_precoding_power}.

Proper power allocation helps controlling the inter-user interference and optimizing the network performance, which often involves advanced optimization techniques \cite{9086497_dowlink_dl_powercontrol}. In \cite{7827017_cellfree_vs_smallcells}, authors show that max-min power control enables cell-free massive MIMO to provide uniformly good service to all users, regardless of their locations. The channel hardening property of cell-free MIMO allows neglecting small-scale fading, causing long-term fading to determine the power controlling time \cite{9086497_dowlink_dl_powercontrol}. Higher time-complexity of optimization-based power control becomes a challenge to meet these time constraints and limits their practical implementation. 

Owing to the universal function approximation property of artificial neural networks (ANNs) \cite{HORNIK1989359_ANN}, DL-based techniques have enabled radio resource allocation with a lower complexity than traditional optimization-based approaches. Several studies have proposed DL-based power control for cellular and cell-free massive MIMO systems \cite{9086497_dowlink_dl_powercontrol, 8645343_mimo_dl_powercontrol, 9022520_dl_cellfree_powercontrol, 8922744_dl_powercontrol, 9104036_mimo_emil}. Most of the existing studies focus on supervised learning approach where a DNN is trained to learn the mapping between the inputs (user locations or channel statistics) and the optimal power allocations obtained by an optimization algorithm. The unsupervised learning algorithm proposed in \cite{8922744_dl_powercontrol} for $K$-user interference channel power control problem eliminates the need of knowing the optimal power allocations during model training. In this study, we are interested in such an unsupervised learning algorithm for cell-free massive MIMO power control which will simplify the data preparation and model training stages. The contributions of the paper are as follows:

\begin{itemize}

    \item In this paper, we consider the max-min rate problem in a cell-free massive MIMO system. We propose, design and implement a DNN to learn user power allocations using channel statistics to achieve max-min user fairness in an unsupervised manner. The method consists of an offline model training stage and an online prediction stage.

    \item In contrast to previous work on supervised learning-based power control, we introduce unsupervised learning for cell-free massive MIMO power control. It does not require the optimal power allocations to be known during model training as in supervised learning, which makes the data preparation and model training simpler, more practical and flexible, since the DNN can be easily retrained in a changing environment over the time.

    \item Furthermore, we introduce an online training stage to improve the performance. The model is customized and fine-tuned in each channel realization during the online implementation in order to improve minimum user rate.
    
    \item Simulation results show that the proposed DNN achieves close performance to the conventional optimization-based max-min power control, with a significantly lower time-complexity. The performance-complexity trade-off of the proposed DL-based approach makes it a potential candidate for practical implementation.

\end{itemize}

\section{System Model}

We consider a cell-free massive MIMO system with $M$ single-antenna APs and $K$ single-antenna users randomly distributed in a $D \times D$ geographic area. The APs are connected to a CPU via backhaul connections. The channel coefficient between $k$th user and $m$th AP is modeled as $g_{mk} = \sqrt{\beta_{mk}} h_{mk}$ \cite{7827017_cellfree_vs_smallcells}. Here, $\beta_{mk}$ is the large-scale fading consisting of pathloss and shadowing, and $h_{mk}\sim \mathcal{CN}(0,1)$ represents the small-scale fading between $k$th user and $m$th AP. The uplink of the network is considered which consists of pilot transmission, channel estimation, and uplink data transmission phases.

\subsection{Pilot Transmission and Channel Estimation}

Initially, all the users undergo a pilot transmission phase in order to estimate the uplink channel coefficients. During this stage, all $K$ users simultaneously transmit their pilot sequences of length $\tau$ symbols to the APs. Let $\sqrt{\tau} \pmb{\phi}_{k} \in \mathbb{C}^{\tau \times 1}$ be the pilot sequence assigned to $k$th user with $ {\parallel \pmb{\phi}_{k} \parallel^{2}} = 1$. The received signal at $m$th AP is then given by

\vspace{-3mm}

\begin{equation}
\label{eq:rx_pilot}
 \textbf{y}_{p,m} = \sqrt{\tau \rho_{p}} \sum_{k=1}^{K} g_{mk} \pmb{\phi}_{k} + \textbf{w}_{p,m},
\end{equation}

\noindent where $\textbf{w}_{p,m} \in \mathbb{C}^{\tau \times 1}$ is the additive noise at $m$th AP with i.i.d $\mathcal{CN}(0,1)$ elements. Then, $m$th AP estimates the channel $g_{mk}, \forall k$ by projecting the received signal $\textbf{y}_{p,m}$ onto pilot sequence $\pmb{\phi}_{k}^H $ as $\tilde{y}_{p,mk} = \pmb{\phi}_{k}^H \textbf{y}_{p,m}$ \cite{7827017_cellfree_vs_smallcells}. Thus, 

\vspace{-5mm}

\begin{equation}
\tilde{y}_{p,mk} = \sqrt{\tau \rho_{p}} \Big(g_{mk}+ \sum_{k'\neq 1}^{K} g_{mk'} \pmb{\phi}_{k}^H \pmb{\phi}_{k'}\Big) + \pmb{\phi}_{k}^H \textbf{w}_{p,m},
\end{equation}

\noindent where the linear minimum mean-squared error (MMSE) estimate of $g_{mk}$ given $\tilde{y}_{p,mk}$ is

\begin{equation}
\label{mmse}
\hat{g}_{mk} = \frac{\mathbb{E} \{ \tilde{y}_{p,mk}^* g_{mk} \}}  {\mathbb{E} \{ \mid \tilde{y}_{p,mk} \mid^2\} } \tilde{y}_{p,mk} = c_{mk} \tilde{y}_{p,mk},
\end{equation}

\noindent where $c_{mk}$ is obtained as \cite{7827017_cellfree_vs_smallcells}

\begin{equation}
c_{mk} = \frac{ \sqrt{\tau \rho_{p}} \beta_{mk}}{\tau \rho_{p} \sum_{k'= 1}^{K} \beta_{mk'} \mid \pmb{\phi}_{k}^H \pmb{\phi}_{k'} \mid^2 + 1}.
\end{equation}

\subsection{Uplink Data Transmission}

After the training phase, the actual uplink data transmission begins where all the users simultaneously send their signals to the APs. Let $x_{k} = \sqrt{\rho \hspace{0.5em} q_{k}} s_k$ be the transmit signal from $k$th user, where $s_k$ is the transmit symbol with $ \mathbb{E} \{\mid s_k \mid^2\} = 1$. Normalized uplink SNR is denoted by $\rho$ and $q_k$ is the power control coefficient of $k$th user, where $0 \leq q_k \leq 1$. The received signal at $m$th AP from all the users is given by

\vspace{-5mm}

\begin{equation}
\label{eq:rx_data}
 {y}_{m} =  \sum_{k=1}^{K} g_{mk} x_k + w_m = \sqrt{\rho}  \sum_{k=1}^{K} g_{mk} \sqrt{q_k} s_k + w_m,
\end{equation}

\noindent where $w_{m}\sim \mathcal{CN}(0,1)$ is the additive noise at $m$th AP. Then match filtering is done at each AP using the locally obtained channel estimate $\hat{g}_{mk}$, and the scaled received signals are sent to the CPU for joint detection. The aggregated received signal $r_k$ (\ref{eq:rx_cpu_basic}) at the CPU is used to detect $s_k$. We assume the large scale fading $\beta_{mk}$ to be known \cite{7827017_cellfree_vs_smallcells}.

\vspace{-4mm}
\begin{align}
\label{eq:rx_cpu_basic}
     r_k = \sqrt{\rho} \sum_{k'= 1}^{K} \sum_{m=1}^{M}  \hat{g}_{mk}^* g_{mk'}  \sqrt{q_{k'}} s_{k'} + \sum_{m=1}^{M}   \hat{g}_{mk}^* w_m. 
\end{align}

\begin{figure*}[ht]
\begin{equation}
\label{eq:user_rate_basic}
R_k^{UP} = \text{log}_2 \Bigg( 1 + \frac{q_k\bigg(\sum_{m=1}^{M} \gamma_{mk}\bigg)^2}
{\sum_{k' \neq k}^{K} q_{k'}\bigg(\sum_{m=1}^{M} \gamma_{mk} \frac{\beta_{mk'}}{\beta_{mk}}\bigg)^2  \mid \pmb{\phi}_{k}^H \pmb{\phi}_{k'} \mid^2 + \sum_{k'=1}^{K} q_{k'} \sum_{m=1}^{M} \gamma_{mk} \beta_{mk'} + \frac{1}{\rho} \sum_{m=1}^{M} \gamma_{mk}} \Bigg).
\end{equation}
\end{figure*}

\subsection{Max-Min User Rate Scheme}
\label{sec:max_min}

In this work, we consider a max-min user fairness scheme where the objective is to maximize the minimum user rate by optimizing the user power allocations. We assume that only the knowledge of channel statistics is used at the CPU when deriving the achievable rate of each user. The uplink rate for the $k$th user can be derived as (\ref{eq:user_rate_basic}) \cite{7827017_cellfree_vs_smallcells}. There, $\gamma_{mk} = \mathbb{E} \{ \mid \hat{g}_{mk} \mid^2 \} = \sqrt{\tau \rho_{p}} \beta_{mk} c_{mk}$. Therefore, we can see that the achievable rate in (\ref{eq:user_rate_basic}) is a function of only the large-scale fading $\beta_{mk}$ and user transmit power coefficients $q_k$, and does not involve instantaneous channel values. Then, the max-min rate problem can be formulated as

\begin{equation}
\begin{aligned}
\label{eq:max-min_basic}
P1:  & \max_{q_k} && \min_{k=1,2,...,K} R_k^{UP}, \\
& \textrm{s.t.} && 0 \leq q_k \leq  1,\quad k=1,2,...,K.\\
\end{aligned}
\end{equation}

In \cite{7827017_cellfree_vs_smallcells}, an algorithm using bisection and solving a sequence
of linear feasibility problems is proposed to solve problem P1. In \cite{8630677_powercontrol}, a less complex algorithm is proposed by reformulating the original problem into a geometric programming (GP) problem and solving it using a convex optimization software to obtain optimum power allocations. Instead of using such an analytical method, we propose a data-driven approach to learn the optimum solutions of the max-min problem.

\section{Deep Learning-based Power Control}
\label{sec:DLpower}

As mentioned earlier, we are interested in an unsupervised learning-based approach which does not require optimal ground truth outputs for model training. For the considered power control problem, a supervised learning approach complicates data preparation and model training due to the complexity of generating ground truth power allocations using an optimization algorithm, especially when $M$ and $K$ are large. In contrast, an unsupervised DNN can be directly fed with inputs in order to learn the optimum solutions minimizing a given loss function during the training process. Such approach is more flexible and adaptable to be practically implemented in a changing wireless communications environment.

\begin{figure}[ht]
\centerline{\includegraphics [width=0.52\textwidth]{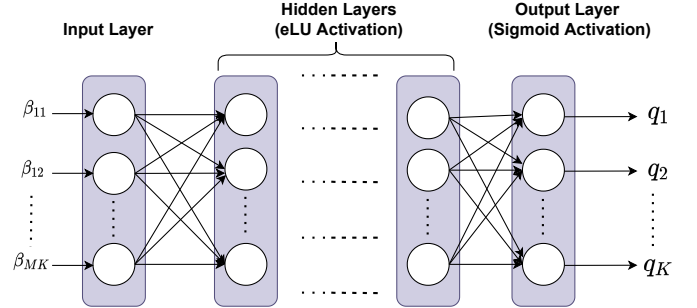}}
\caption{Model layout of power control DNN.}
\label{fig:powerctrlnet}
\end{figure}

We propose a feedforward DNN as illustrated in Fig. \ref{fig:powerctrlnet} in order to address the power control problem P1 (\ref{eq:max-min_basic}). The network consists of $L+1$ layers that are sequentially connected to produce the mapping $f(\textbf{x}_{0};\mbox{\boldmath$\theta$}) : \mathbb{R}^{N_{0} \times 1} \mapsto \mathbb{R}^{N_{L} \times 1} $ of an input vector $\textbf{x}_{0} \in \mathbb{R}^{N_{0} \times 1}$ to an output vector $\textbf{x}_{L} \in \mathbb{R}^{N_{L} \times 1}$ through $L$ iterative processing steps

\vspace{-2mm}

\begin{equation}
    \textbf{x}_{l} = f_{l}(\textbf{x}_{l-1};\theta_{l}), \hspace{4ex} l=1,2,...,L.
\end{equation}
\label{eq:mapping}

\vspace{-3mm}

Here $f_{l}(\textbf{x}_{l-1};\theta_{l}) : \mathbb{R}^{N_{l-1} \times 1} \mapsto \mathbb{R}^{N_{l} \times 1} $ is the mapping performed by the \textit{l}th layer which depends on the output vector $\textbf{x}_{l-1}$ from the previous layer and the set of learnable parameters $\theta_{l}$ in the \textit{l}th layer. The set $\mbox{\boldmath$\theta$} = \{\theta_{1}, \theta_{2},..,\theta_{L} \}$ denotes the set of all the parameters of the network which are learnt through model training.

Input to the network is the large-scale channel coefficients $\beta_{mk}$ of all the APs and users, aligned as a column vector denoted by $\boldsymbol{\beta}$ with dimension $N_0 = MK$. Thus, the DNN input is $\textbf{x}_{0}=\boldsymbol{\beta} \in \mathbb{R}^{MK \times 1}$. The model outputs the estimated power allocation vector $\textbf{q} = [q_1, q_2,...,q_K]^T \in \mathbb{A}^{K \times 1}$, where $\mathbb{A} = \{ a\in \mathbb{R}: 0 \leq a \leq 1 \}$. We have implemented a fully connected neural network with \textit{Dense} layers where $f_{l}(\textbf{x}_{l-1};\theta_{l})$ has the form

\vspace{-4mm}

\begin{equation}
\label{eq:dense}
    f_{l}(\textbf{x}_{l-1};\theta_{l}) = \sigma(\textbf{W}_{l} \textbf{x}_{l-1} + \textbf{b}_{l}), \hspace{3ex} l=1,2,...,L,
\end{equation}

\noindent where $\textbf{W}_{l} \in \mathbb{R}^{N_{l}\times N_{l-1}}$ is the weight matrix, $\textbf{b}_{l} \in \mathbb{R}^{N_{l} \times 1}$ is the bias vector. Then, the set of learnable parameters is $\theta_{l} = \{\textbf{W}_{l},\textbf{b}_{l}\}$. In (\ref{eq:dense}), $\sigma(\cdot)$ is a called an \textit{activation function} such as \textit{ReLU}, \textit{eLU}, \textit{Sigmoid} etc. which introduces non-linearity to the network. For the $L$ hidden layers in the model, we have used the \textit{eLU} (exponential linear unit) activation function. For the output layer, \textit{Sigmoid} activation function is used to guarantee that the outputs are in the range $[0,1]$ adhering to the transmit power constraints $0 \leq q_k \leq 1, \forall k $ users.

We implemented a DNN with 4 layers ($L=3$) which has $\{MK, K, M, K\}$ number of neurons in each layer and $\{\textit{eLU}, \textit{eLU}, \textit{eLU}, \textit{Sigmoid}\}$ activations respectively. This DNN has a considerably simpler structure than the DNN proposed in \cite{9022520_dl_cellfree_powercontrol} in terms of network dimensions. Therefore, it has a lower training complexity and can produce outputs with a lower online complexity.

Given that the goal of problem P1 in (\ref{eq:max-min_basic}) is to maximize the minimum user rate, we apply following loss function for model training as 

\vspace{-2mm}

\begin{equation}
\label{eq:lossfunction}
    loss = - \mathbb{E}_{\boldsymbol{\beta}}[R(\boldsymbol{\beta},\boldsymbol{\theta})_{min}],
\end{equation}

\noindent where $\boldsymbol{\theta}$ denotes the set of trainable parameters in the model. There, $R(\boldsymbol{\beta},\boldsymbol{\theta})_{min} = \min_{k=1,2,...,K} R(\boldsymbol{\beta},\boldsymbol{\theta})_k^{UP}$ is the minimum user rate among all the $K$ users for a given channel realization with large-scale fading of $\boldsymbol{\beta}$ and given $\boldsymbol{\theta}$. For each user $k$, $R(\boldsymbol{\beta},\boldsymbol{\theta})_k^{UP}$ is calculated from (\ref{eq:user_rate_basic}) using $\boldsymbol{\beta}$ and $\textbf{q}(\boldsymbol{\theta})$ where $\textbf{q}(\boldsymbol{\theta})$ is the output from DNN for given $\boldsymbol{\theta}$. 

This loss function is differentiable with $\boldsymbol{\theta}$ which allows training the network via stochastic gradient descent (SGD). We adopt mini-batch gradient descent approach to reduce the complexity of the SGD. In each iteration of the training, a set of channel realizations are generated from its distribution. Thus, the training loss is approximated as

\vspace{-2mm}

\begin{equation}
\label{eq:lossfunction_minibatch}
    loss \approx - \frac{1}{\mid \mathcal{B} \mid} \sum_{\boldsymbol{\beta} \in \mathcal{B}} [R(\boldsymbol{\beta},\boldsymbol{\theta})_{min}],
\end{equation}

\noindent where $\mathcal{B}$ denotes the set of channel realizations in each iteration and $\mid \mathcal{B} \mid$ is the mini-batch size. Thus, during training, the model learns parameters $\boldsymbol{\theta}$ to minimize the loss given in (\ref{eq:lossfunction_minibatch}) which maximizes the minimum user rate as expected.

\section{Simulations and Results}

In this section we present the numerical simulations to evaluate the performance of the proposed DL-based max-min fairness scheme in comparison with existing optimization-based techniques.

\subsection{Simulation Parameters}

We consider a cell-free MIMO system in a simulation area of $1 \times 1 \hspace{2mm} km^2$ with different number of AP and user configurations. This square area is wrapped around at the edges to avoid boundary effects and to emulate a cell-free network with an infinite area \cite{7827017_cellfree_vs_smallcells}. We refer to \cite{7827017_cellfree_vs_smallcells} and Table \ref{tab:simulation} for more details about simulation parameters.

The large-scale fading coefficient $\beta_{mk}$ from $k$th user to $m$th AP is given by \cite{7827017_cellfree_vs_smallcells}

\vspace{-2mm}

\begin{equation}
    \beta_{mk} = PL_{mk} 10^{\frac{\sigma_{sh} z_{mk}}{10}},
    \label{eq:pathloss_and_shadow}
\end{equation}

\noindent where $PL_{mk}$ is the pathloss from $k$th user to $m$th AP, calculated using the three-slope model used in \cite{7827017_cellfree_vs_smallcells} with parameters as in Table \ref{tab:simulation}. The shadow fading with standard deviation $\sigma_{sh}$, and $z_{mk}\sim \mathcal{N}(0,1)$ is denoted by $10^{\frac{\sigma_{sh} z_{mk}}{10}}$.

\begin{table}[ht]
\caption{Simulation parameters}
\begin{center}
\begin{tabular}{|c|c|}
\hline
\textbf{Parameter} & \textbf{Value} \\ \hline 
Carrier frequency ($f$) & 1.9 GHz \\ \hline
Simulation area length ($D$) & 1 km \\ \hline
AP antenna height ($h_{AP}$) & 15 m \\ \hline
User antenna height ($h_{u}$) & 1.65 m \\ \hline
$d_0$, $d_1$ & 10, 50 m \\ \hline
Bandwidth ($B$) & 20 MHz \\ \hline
Noise figure & 9 dB \\ \hline
$\sigma_{sh}$ & 8 dB \\ \hline
Pilot and data transmit powers ($\bar{\rho}_p$, $\bar{{\rho}}$) & 100, 100 mW \\ \hline
\end{tabular}
\end{center}
\label{tab:simulation}
\end{table}

When evaluating the spectral efficiencies, per-user net throughputs are considered as follows accounting for the channel estimation overhead as well.

\vspace{-1mm}

\begin{equation}
    R_k^{net} = B \frac{(1 - \tau/\tau_c)}{2} R_k^{UP},
\end{equation}

\noindent where $R_k^{UP}$ is the per-user rate in (\ref{eq:user_rate_basic}), $B$ is the spectral bandwidth,  and $\tau_c$ is the coherence interval in samples. We have used $\tau_c=200$ corresponding to a coherence bandwidth of 200 kHz and a coherence time of 1 ms. Orthogonal pilot assignment is considered in all simulations where $\tau = K$ so that each user is assigned with a unique orthogonal pilot sequence.

\subsection{Simulation Setup}

DNN model implementation, training and testing is done in TensorFlow \cite{tensorflow}. Optimization-based baseline implementation is done in Matlab using CVX convex optimization software package \cite{cvx}, \cite{gb08}. Both implementations are done on the same platform with a 4-core Intel(R) Core(TM) i5-8250U CPU with 1.6 GHz frequency.

We used three different datasets for DNN model training, validation and testing, consisting of $10^5$, 1000 and 1000 different samples respectively. For each sample, different AP and user distributions were considered with randomly generated large-scale fading channel coefficients. Input to the DNN is normalized using the training dataset mean and variance. The network is trained for 10000 iterations, using mini-batch gradient descent along with the ADAM optimizer with learning rate 0.01. In each iteration, a random mini-batch of size 100 is selected from the training dataset. In every 50 iterations, validation dataset is used to evaluate the model, where the model parameters corresponding to the minimum validation loss are preserved along the training. After training, performance is evaluated for the test dataset where the trained model is used to produce the power allocations for the test dataset and to calculate per-user rates using (\ref{eq:user_rate_basic}).

The GP optimization algorithm proposed in \cite{8630677_powercontrol} is used as the baseline for performance comparison and solved using CVX. Maximum power transmission where all users transmit with full power (i.e. $q_k=1, \forall k$) is also considered. CVX-based max-min power control results and maximum power transmission results for the test dataset are denoted as ``baseline" and ``maximum-power" respectively, in the results section.

\begin{figure}[ht]
\centerline{\includegraphics [width=0.5\textwidth]{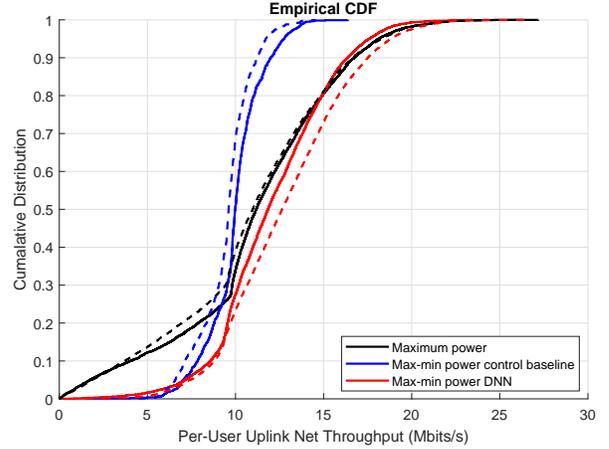}}
\caption{Cumulative distribution of the per-user net throughput for $M=30, K=5$ (solid lines) and $M=50, K=10$ (dashed lines).}
\label{fig:M30,50_userrate_orthogonal}
\end{figure}

\begin{figure}[ht]
\centerline{\includegraphics [width=0.5\textwidth]{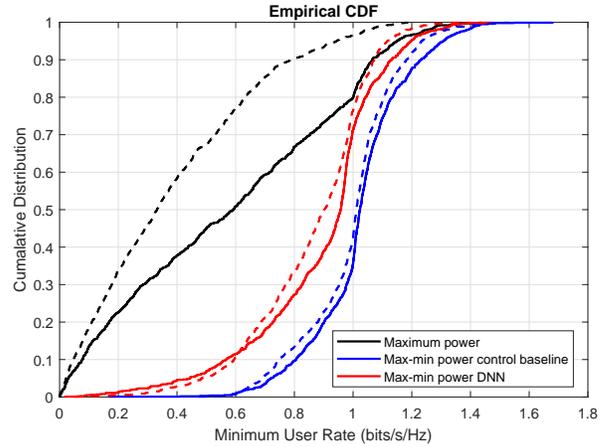}}
\caption{Cumulative distribution of the minimum user rate for $M=30, K=5$ (solid lines) and $M=50, K=10$ (dashed lines).}
\label{fig:M30,50_minrate_orthogonal}
\end{figure}

\subsection{Results and Discussion}

\begin{figure}[ht]
\centerline{\includegraphics [width=0.53\textwidth]{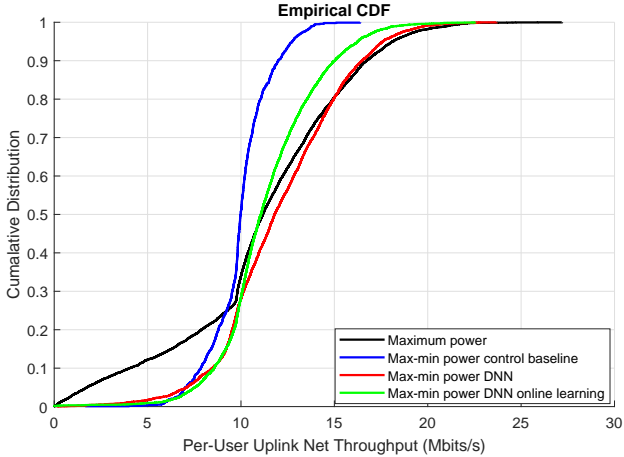}}
\caption{Cumulative distribution of the per-user net throughput for $M=30, K=5$, with online training.}
\label{fig:M30,5_userrate_onlinetraining}
\end{figure}

\begin{figure}[ht]
\centerline{\includegraphics [width=0.53\textwidth]{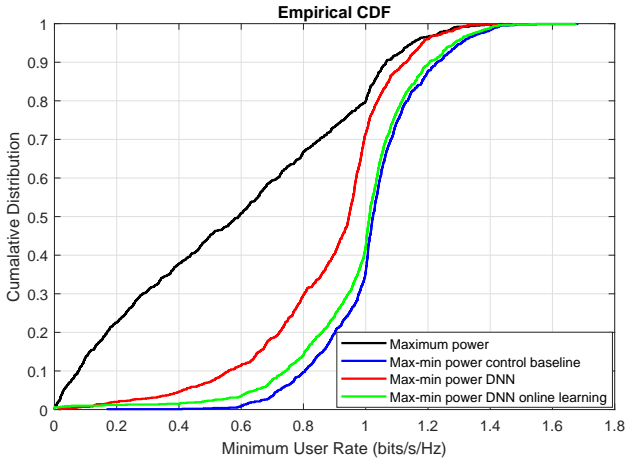}}
\caption{Cumulative distribution of the minimum user rate for $M=30, K=5$ with online training.}
\label{fig:M30,5_minrate_onlinetraining}
\end{figure}

For performance comparison, we consider two scenarios with $M=30, K=5$ and $M=50, K=10$ randomly distributed over the $1 \times 1$ km\textsuperscript{2} simulation area. Obtained cumulative distribution curves are presented in Fig. \ref{fig:M30,50_userrate_orthogonal}. The DNN has close performance to the baseline in the lower net throughput range. Both the baseline and DNN have almost same 95\% likely per-user net throughput in both network configurations. However, the difference in the baseline and DNN performance suggests that the DNN may have learnt a sub-optimal power allocation scheme. In order to assess how well the DNN has achieved the desired objective of max-min user fairness, we have also evaluated the minimum user rate performance, which is shown in Fig. \ref{fig:M30,50_minrate_orthogonal}. Even though the DNN has a lower minimum user rate performance than the optimal baseline solution, it has a significant improvement over the maximum power scenario.

Furthermore, exploiting the unsupervised learning capability of the model, we have introduced online training to improve the performance of the DNN. There, the originally trained model (with the training set) is retrained for 100 iterations with learning rate 0.01 for each input sample in the test set. Performing online training allows further customization and fine-tuning of model parameters based on large-scale channel inputs in each channel realization, further improving the minimum user rate. Fig. \ref{fig:M30,5_userrate_onlinetraining} and Fig. \ref{fig:M30,5_minrate_onlinetraining} show the improved results for the per-user rate and minimum user rate cumulative distributions with online training. It can be seen that the minimum user rates are significantly improved with online training, resulting in worst case user performance much closer to the optimal. Obtained average minimum user rates over the training and test datasets for baseline and DNN implementations in all simulation scenarios are summarized in Table \ref{tab:min_rate}. 

\begin{figure}[ht]
\centerline{\includegraphics [width=0.53\textwidth]{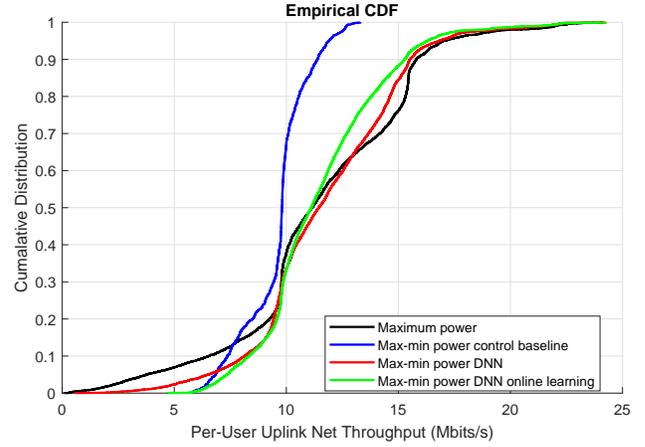}}
\caption{Cumulative distribution of the per-user net throughput for $M=30, K=5$, with fixed APs and moving users.}
\label{fig:M30,5_userrate_fixedAP,movingUE}
\end{figure}

\begin{figure}[ht]
\centerline{\includegraphics [width=0.53\textwidth]{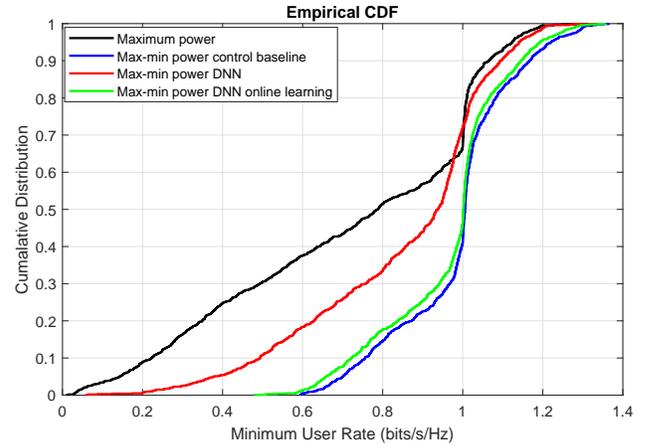}}
\caption{Cumulative distribution of the minimum user rate for $M=30, K=5$, with fixed APs and moving users.}
\label{fig:M30,5_minrate_fixedAP,movingUE}
\end{figure}

We also evaluated the performance of the proposed DNN model with a fixed AP setup and moving user scenario with $M=30, K=5$. The same DNN architecture and training process is used, but with a different data set. Here, we consider that the APs are located in a regular grid. Each user starts with a random initial position which is uniformly distributed in the coverage area, and moves in a random direction (left, right, up and down) in a random speed uniformly distributed between 0 and 20 m/s. The moving speed and direction are changed in every 5s. If a user reaches the boundary of the coverage area, then the direction is reversed so that it remains inside the area. A dataset of 12000 samples corresponding to 12000s is generated in this manner, and divided into 10000, 1000 and 1000 samples to be used as training, validation and test sets respectively. Fig. \ref{fig:M30,5_userrate_fixedAP,movingUE} and Fig. \ref{fig:M30,5_minrate_fixedAP,movingUE} show the per-user rate and minimum user rate cumulative distributions obtained for this setup. DNN implementations show better results than in the earlier random AP/user scenario, achieving near-optimal minimum user rate performance with online training. Note that here the reported performance was achieved only using a training set of 10000 samples and spending less than 10 minutes for offline model training. This shows the potential of the proposed unsupervised DNN for quick and easy deployment in a practical setup due to its low complexity offline training and online processing, and the comparable performance.

\setlength{\tabcolsep}{7pt}
\setlength{\extrarowheight}{2pt}

\begin{table}[h]
\caption{Average minimum user rate comparison for DNN and baseline.}
\begin{center}
\begin{tabular}{|l|l|l|l|}
\hline
\textbf{System} & \textbf{Baseline} & \textbf{DNN}  &  \textbf{DNN with} \\ 
\textbf{Setup} & & & \textbf{Online Training} \\
\hline
$M=30$ & Train: 1.0218 & Train: 0.8747 & Train: 0.8747\\
$K=5$ & Test: 1.0221 & Test: 0.8653 & Test: 0.9817 \\
\hline
$M=50$ & Train: 0.9897 & Train: 0.7966 & Train: 0.7966\\
$K=10$ & Test: 0.9940 & Test: 0.7816 & Test: 0.9199 \\
\hline
$M=30$ & Train: 0.9889 & Train: 0.8842  & Train: 0.8842 \\
$K=5$ & Test: 0.9854 & Test: 0.8473 & Test: 0.9664 \\
Moving users & &  & \\
\hline
\end{tabular}
\end{center}
\label{tab:min_rate}
\end{table}

\section{Complexity Analysis}

Here we compare the computational complexity of the baseline methods and the proposed DNN implementation for solving problem P1. Bisection-based algorithm proposed in \cite{7827017_cellfree_vs_smallcells} has a complexity of $\text{log}_2(\frac{t_{max}-t_{min}}{\epsilon}) \mathcal{O}(K^4)$ \cite{8630677_powercontrol}. The GP-based low complexity approach proposed in \cite{8630677_powercontrol} has $\mathcal{O}(K^{7/2})$ complexity. The newly proposed DNN for approximating solutions for the problem P1 has a complexity of $\mathcal{O}(K^2M)$ considering the dimensions of the proposed DNN model.

The recorded CPU timing for CVX solver and the DNN with and without online training, to produce outputs for 100 channel realizations for $M=30, K=5$ are 46.47s, 10.61s and 0.12s respectively. For $M=50, K=10$, respective computational times are 78.95s, 12.87s and 0.17s. Thus, the DNN is around 400 times faster than the baseline. Furthermore, DNN with online training which has near-optimal rate performance is also around 4-6 times faster than the baseline. This computational complexity of DNNs can be significantly reduced by GPU aided parallel processing implementations which are often used in DL implementations. Furthermore, it is apparent that the fast processing of DNN implementations become more significant with increasing network dimensions $M$ and $K$.

\section{Conclusion}

In this study, we have proposed an unsupervised DL-based algorithm for max-min power control for the uplink of a cell-free massive MIMO system. The proposed DNN produces sub-optimal power allocations resulting in close per-user net throughput performance compared to the GP-based optimal solution. Performing online training to customize the learnt model parameters in each channel realization to further improve the max-min performance results in near-optimal per-user and min-user rate performance at the expense of processing complexity. Nevertheless, the proposed DNN implementations are much less computationally complex than the GP-based optimization algorithm, specially for larger AP and user configurations. Furthermore, the proposed unsupervised learning approach has a lower training complexity than the supervised learning implementation in \cite{9022520_dl_cellfree_powercontrol}, and also has a much lower online complexity due to its simpler network structure compared to \cite{9022520_dl_cellfree_powercontrol}.

While this is the first time unsupervised learning-based power control is implemented for cell-free massive MIMO, it should be noted that we have analyzed the performance for a fairly a simpler network configuration with less number of APs and users than in a practical network setup. Therefore, further investigations need to be done to understand the best DNN architectures and hyper-parameters for the proposed unsupervised learning approach to get better results in such complex setups. Nevertheless, DL-based power control in cell-free massive MIMO has research potential, especially when considering complex scenarios such as joint AP selection/user assignment and power control where conventional approaches might be sub-optimal.

\bibliography{ref}
\bibliographystyle{IEEEtran}

\vspace{12pt}

\end{document}